\documentclass{PoS}

\title{PDF uncertainties in the determination of the  $W$ boson mass
and of the effective lepton mixing angle at the LHC}

\ShortTitle{PDF uncertainties in $M_W$  at the LHC}

\author{\speaker{Juan Rojo}\thanks{This work is supported by a Marie Curie 
Intra--European Fellowship of the European Community's 7th Framework Programme under contract number PIEF-GA-2010-272515. }\\
        PH Department, TH Unit, CERN, CH-1211 Geneva 23, Switzerland\\
        E-mail: \email{juan.rojo@cern.ch}}

\abstract{The precision measurement of the $W$ boson mass allows to
perform stringent consistency tests of the Standard Model by means
of global electroweak fits. 
The accurate determination of the $W$ boson mass is one of the legacy results
of the Tevatron, where the experimental accuracy is such that $M_W$ is
now limited by theoretical uncertainties related to the parton
distributions of the proton.
In this contribution, we show how to quantify the impact of PDF uncertainties in the measurement of $M_W$
at the Tevatron and the LHC by means of  a template
method, and study both the use of the $W$ transverse mass and
the lepton $p_T$ kinematical distributions to generate these
templates.
 We also present preliminary
results on the quantification of the PDF uncertainties in the
determination of the effective lepton mixing angle at the LHC, 
based on the same template method as for the $W$ mass
determination.}

\FullConference{XXI International Workshop on Deep-Inelastic Scattering and Related Subject -DIS2013,\\
		22-26 April 2013\\
		Marseilles,France}

\begin{document}

\paragraph{PDF uncertainties in the determination of $M_W$ at hadron colliders.}

The $W$ boson mass has been accurately measured at the Tevatron~\cite{Group:2012gb}, where the latest combination gives a total uncertainty of about 16 MeV.
PDFs are the dominant theory uncertainty in this measurement.
The precision measurement of the $W$ mass allows to
perform stringent consistency test of the Standard Model by means
of global electroweak fits~\cite{Baak:2012kk}.
With the same motivation, it is of utmost importance 
to provide a precision determination of $M_W$ also at the LHC,
and to assess the feasibility of this goal
one needs first to quantify the role of PDF uncertainties, and
to provide new ideas to reduce their impact in the final 
$M_W$ measurement.

In order to determine $\Delta M_W$ due to PDF uncertainties, a template
method was followed in Ref.~\cite{Bozzi:2011ww}, an strategy
similar to that used by the Tevatron collaborations.
 In a first step, we generate pseudo-data for the
$W$ transverse mass distribution, for various PDF sets and their corresponding
error sets: CTEQ6.6~\cite{Nadolsky:2008zw}, MSTW08~\cite{Martin:2009iq} 
and NNPDF2.1~\cite{Ball:2011mu}. 
Then, with the central
CTEQ6.6 as input, we generate a large number of templates varying
$M_W$ from the reference value in a suitable range. 
The shift in $M_W$ corresponding to each PDF error set is determined
by the template which leads to a better $\chi^2$ agreement with the
corresponding pseudo-data.
An essential ingredient of our approach is to normalize the templates
to the total integral of the distribution in the fit
region, since this way PDF
uncertainties are substantially reduced without losing
sensitivity to the value of $M_W$.

\begin{figure}[h]
\centering
\includegraphics[scale=0.22]{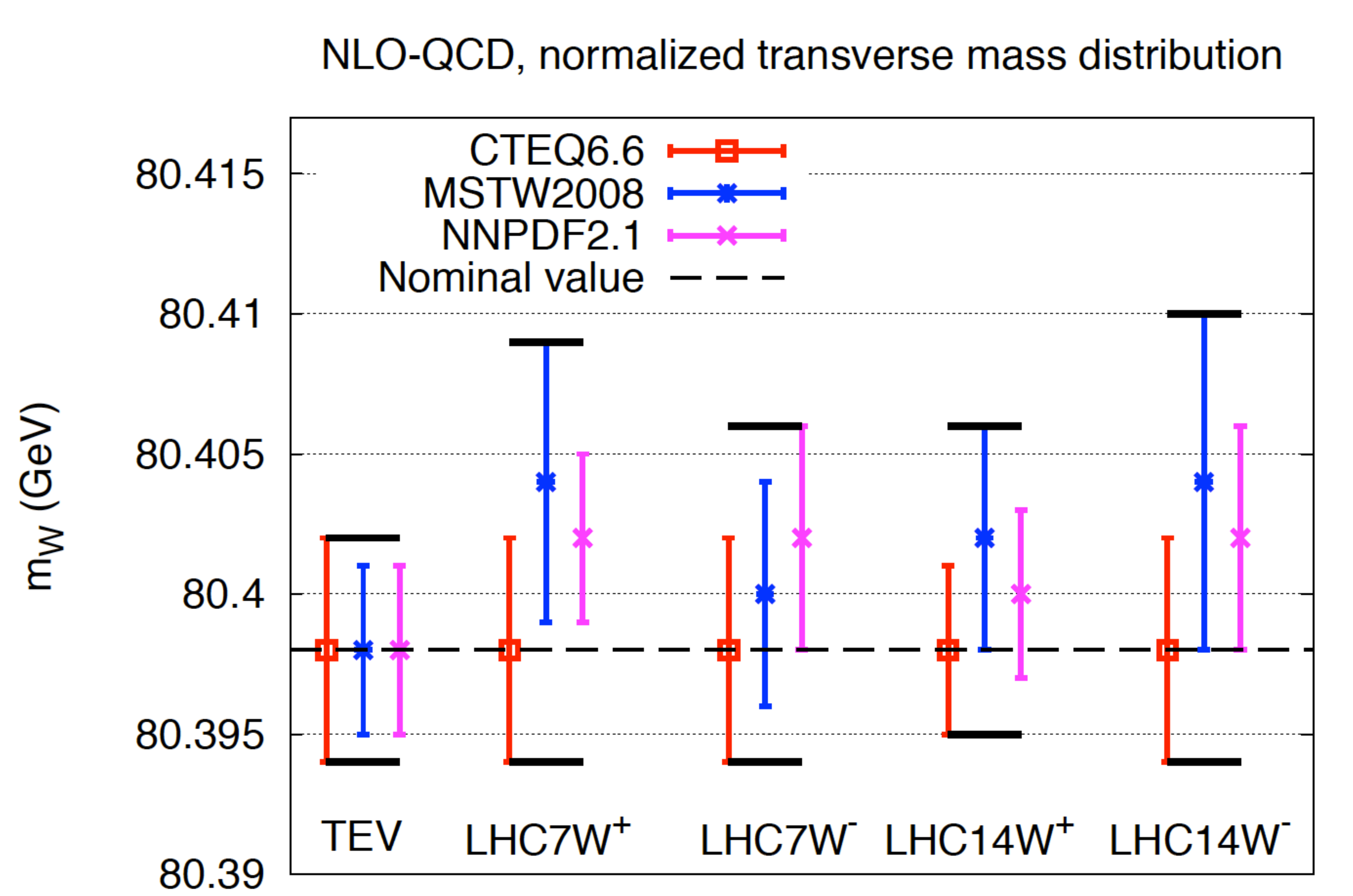}
\caption{\small  PDF uncertainties in the $W$ boson
determination from the transverse mass distribution at the
Tevatron and the LHC. The templates for different $M_W$ have been
generated with the central CTEQ6.6 set.
}
\label{fig:wmass}
\end{figure}

Templates were generated with {\tt HORACE}~\cite{CarloniCalame:2007cd} 
at LO and with
 {\tt DYNNLO}~\cite{Catani:2010en} at NLO.
Our results are summarized in Fig.~\ref{fig:wmass}. 
A conservative estimate is that PDF uncertainties in $M_W$
fits at the LHC will not be larger than 20 MeV.
Note that these results where obtained with PDF sets without
LHC data, and that PDF uncertainties are likely to be reduced
in PDF sets which include the LHC $W$ and $Z$ production data
such as NNPDF2.3~\cite{Ball:2012cx}.
Our results do not support previous studies~\cite{Krasny:2010vd}, 
which claimed
that achieving a 10 MeV accuracy on $M_W$ at the LHC was not
feasible.

\begin{figure}[h]
\centering
\includegraphics[scale=0.37]{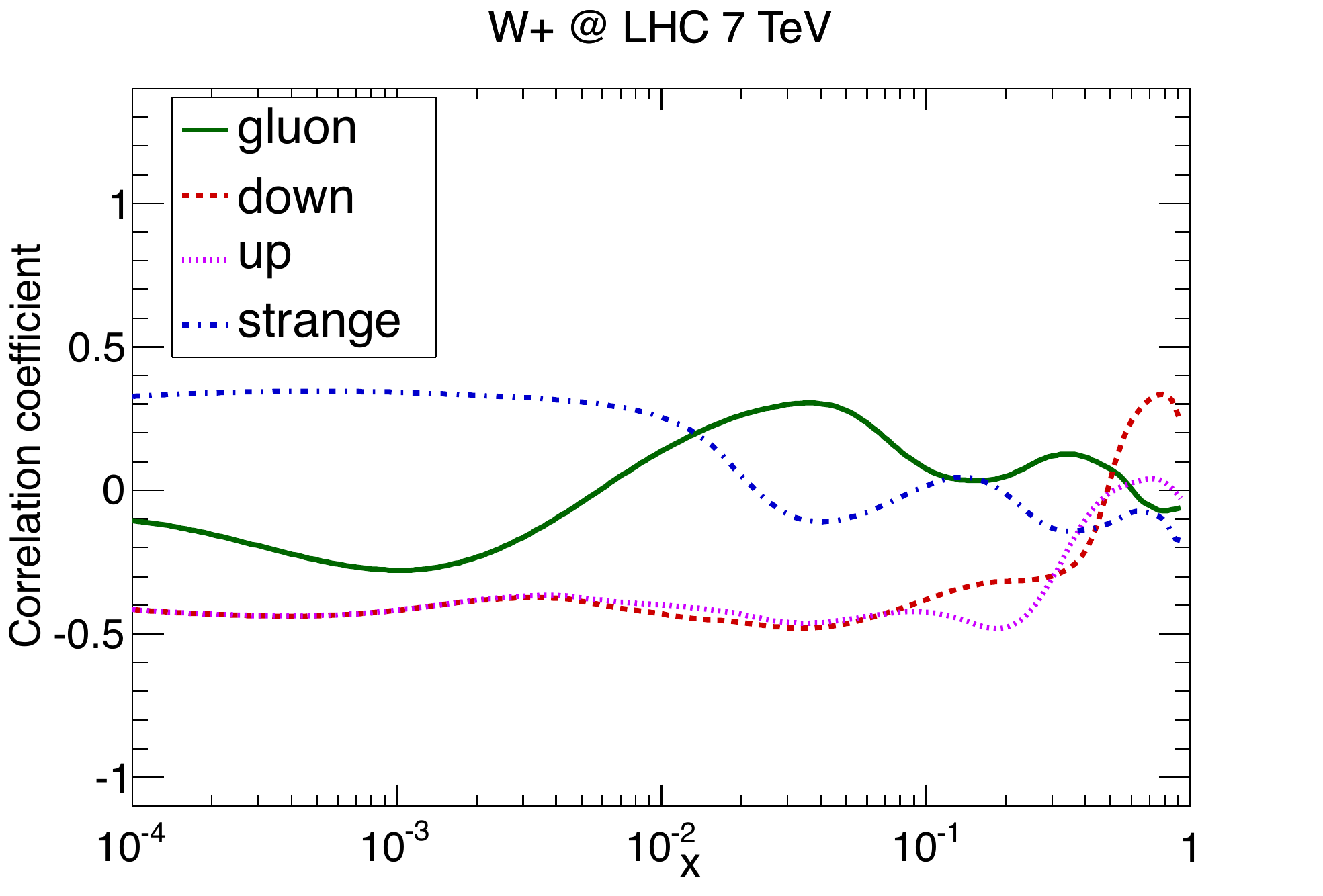}
\includegraphics[scale=0.37]{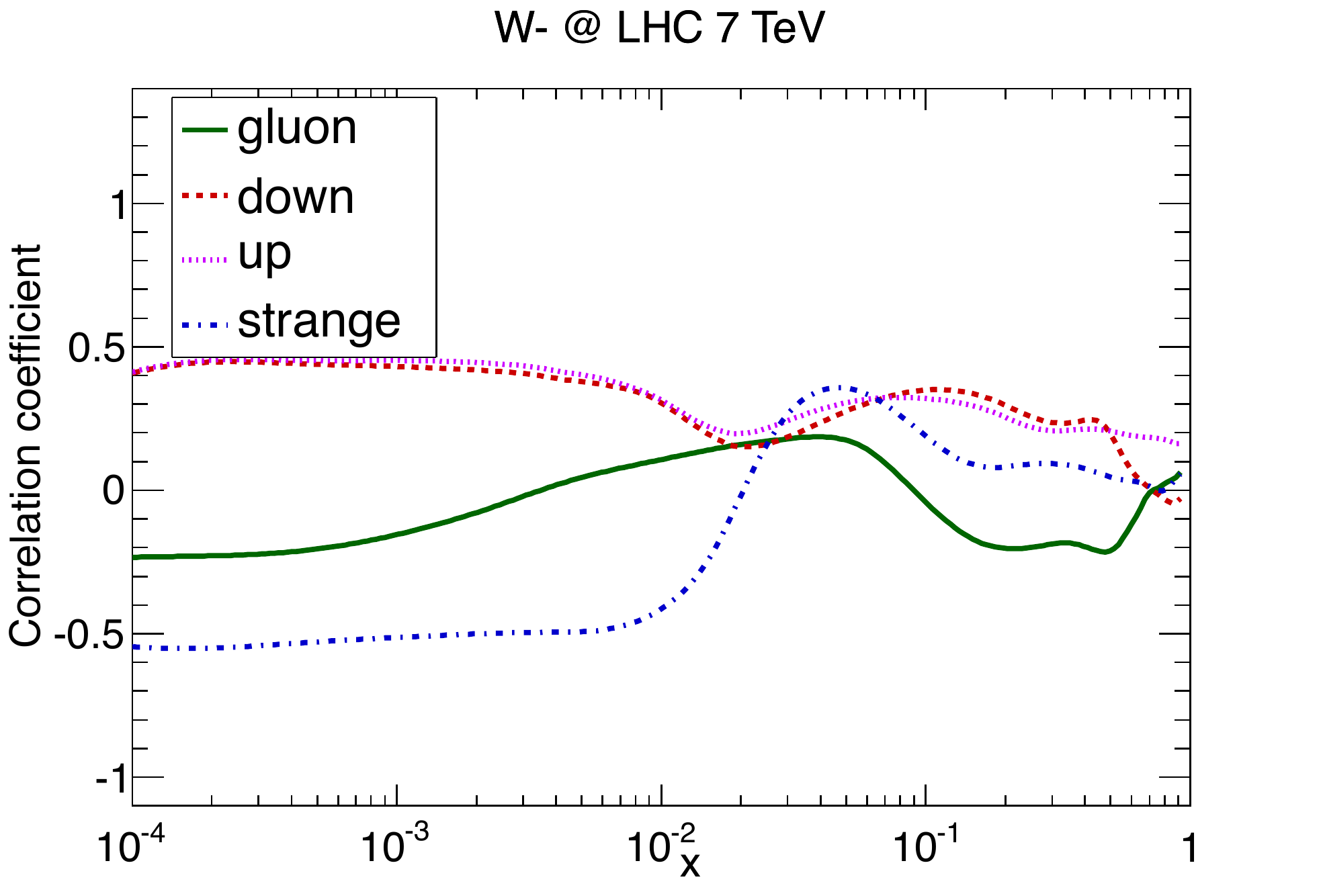}
\caption{\small  Correlations between different PDF flavours
and the $M_W$ determination at LHC 7 TeV, as a function of
Bjorken-$x$. A large absolute value of the correlation coefficient
indicates a strong sensitivity to a particular PDF combination.
}
\label{fig:correlations}
\end{figure}

In order to determine if a particular PDF combination is responsible
for PDF uncertainties in $M_W$, it is useful to compute the
correlations~\cite{Demartin:2010er} 
between the $N_{\rm rep}=100$ PDF replicas of NNPDF2.1 and the 100
determinations of $M_w$ obtained from the template fits for each replica.
The results are shown in Fig.~\ref{fig:correlations}. It is clear that
there is not a particular range of  Bjorken-$x$ or a particular quark flavor
that dominates the $M_W$ measurement. 
The sensitivity to the gluon
is smaller to that to quarks, consistent with the expectation
that the $W$ transverse mass distribution is stable against NLO and higher
order corrections.
This implies that to improve PDF uncertainties in the
$M_W$ measurement new data constraining quarks in a broad
$x$ range would be needed.

\paragraph{ $M_W$ determination from the lepton $p_T$ distribution.}
The previous results were based on the determination of $M_W$
from the $W$ transverse mass distribution. Another observable
that has been
used at the Tevatron is the lepton $p_T$ distribution.
Below we report
preliminary work towards the extension of the results of~\cite{Bozzi:2011ww}
to template fits of the lepton transverse momentum~\cite{prep1}. 

As opposed to the transverse mass distribution, the lepton $p_T$ is
substantially modified by higher order corrections, given its 
strong correlation with the $W$ $p_T$ which vanishes at the Born
level.
 For this distribution the use of resummed calculations for the
$W$ boson $p_T$ is required, either using analytical $p_T$ resummations or
NLO calculations matched to parton showers.

\begin{figure}[h]
\centering
\includegraphics[scale=0.15]{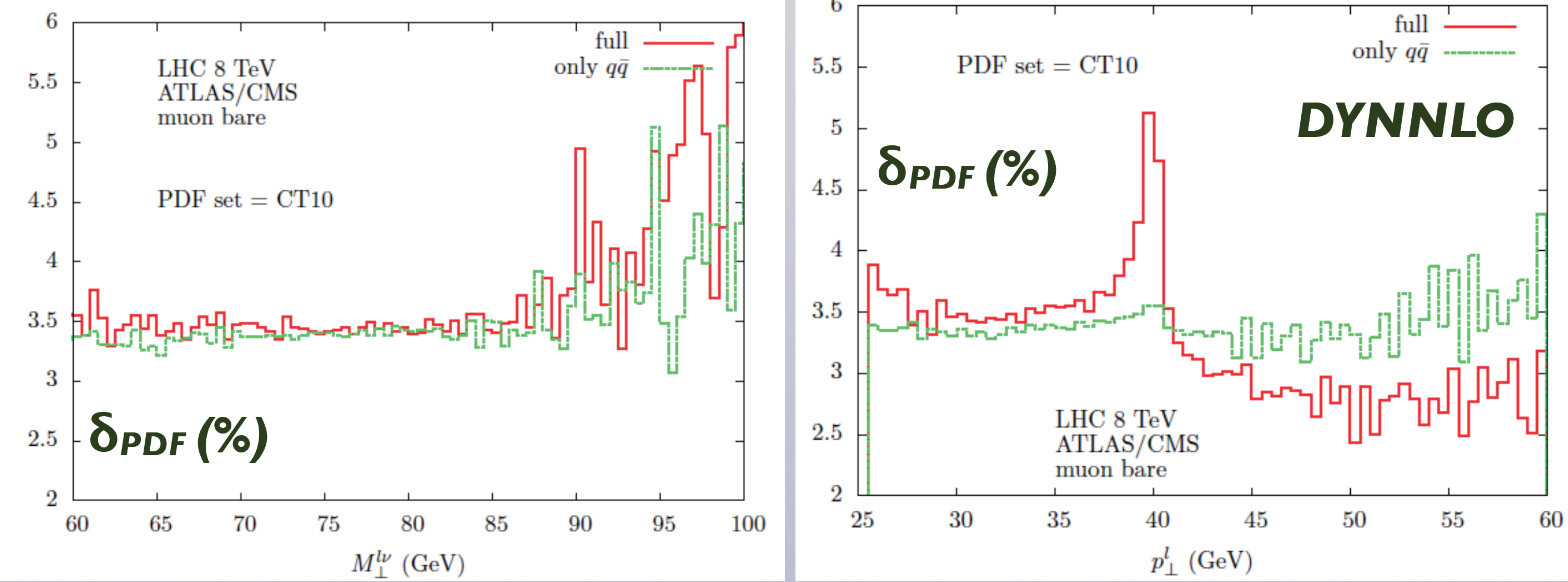}
\caption{\small  The total relative PDF
uncertainty and the separate contribution of quark-antiquark
diagrams  for the transverse
mass (left plot) and the lepton $p_T$ (right plot) distributions,
computed at NLO with {\tt DYNNLO}.
}
\label{fig:pt}
\end{figure}

The relevance of NLO corrections implies that the gluon PDF
will be a more important contribution to the uncertainty in
$M_W$ than in the previous case. In order to confirm this, in
Fig.~\ref{fig:pt} we show the contribution of quark-antiquark
terms to the total PDF uncertainty in the transverse mass and lepton
$p_T$ distributions, computed at NLO with {\tt DYNNLO}.
 Is clear that for the lepton $p_T$ distribution the
$qg$ contributions are substantial, in particular
near the Jacobian peak.
Therefore, a dedicated analysis strategy should be pursued
in order to limit as much as possible the contribution to
$\Delta M_W$ due to the gluon PDF, such as taking ratios
of $W$ over $Z$ distributions.

\paragraph{PDF uncertainties in the determination of $\sin^2\theta^l_{\rm eff}$ at the LHC.}

The same template methods can be used to study the role of 
PDF uncertainties in the determination of the effective lepton
mixing angle, $\sin^2\theta^l_{\rm eff}$, at the LHC~\cite{prep2}.
Currently, $\sin^2\theta^l_{\rm eff}$ 
 is known accurately from global fits to LEP data. 
 However, there is a some degree of tension between different 
inputs to the global electroweak
fits.
 In particular, some of the most precise determinations 
of this quantity are inconsistent. 
Therefore, an additional
independent determination of $\sin^2\theta^l_{\rm eff}$ at the
LHC is potentially relevant.
 However, the LHC determination will
only be competitive with the LEP results if PDF uncertainties
do not spoil the accuracy of the measurement.

Following the same approach as for $M_W$, we have generated
pseudo-data and templates for the forward-backward asymmetry
in $\gamma^*/Z$ production at the LHC 7 TeV, using {\tt HORACE} at
LO. 
The templates have been generated varying the value of 
 $\sin^2\theta^l_{\rm eff}$ from the nominal value. 
The binning of the templates have been optimized in order to have
similar statistical uncertainties in the various bins, in the range 
$60 \le M_{ll} \le 120$ GeV.
The kinematical cuts and acceptances have been selected to
simulate both the typical selection cuts from ATLAS and CMS,
and then for LHCb.

\begin{figure}[h]
\centering
\includegraphics[scale=0.70]{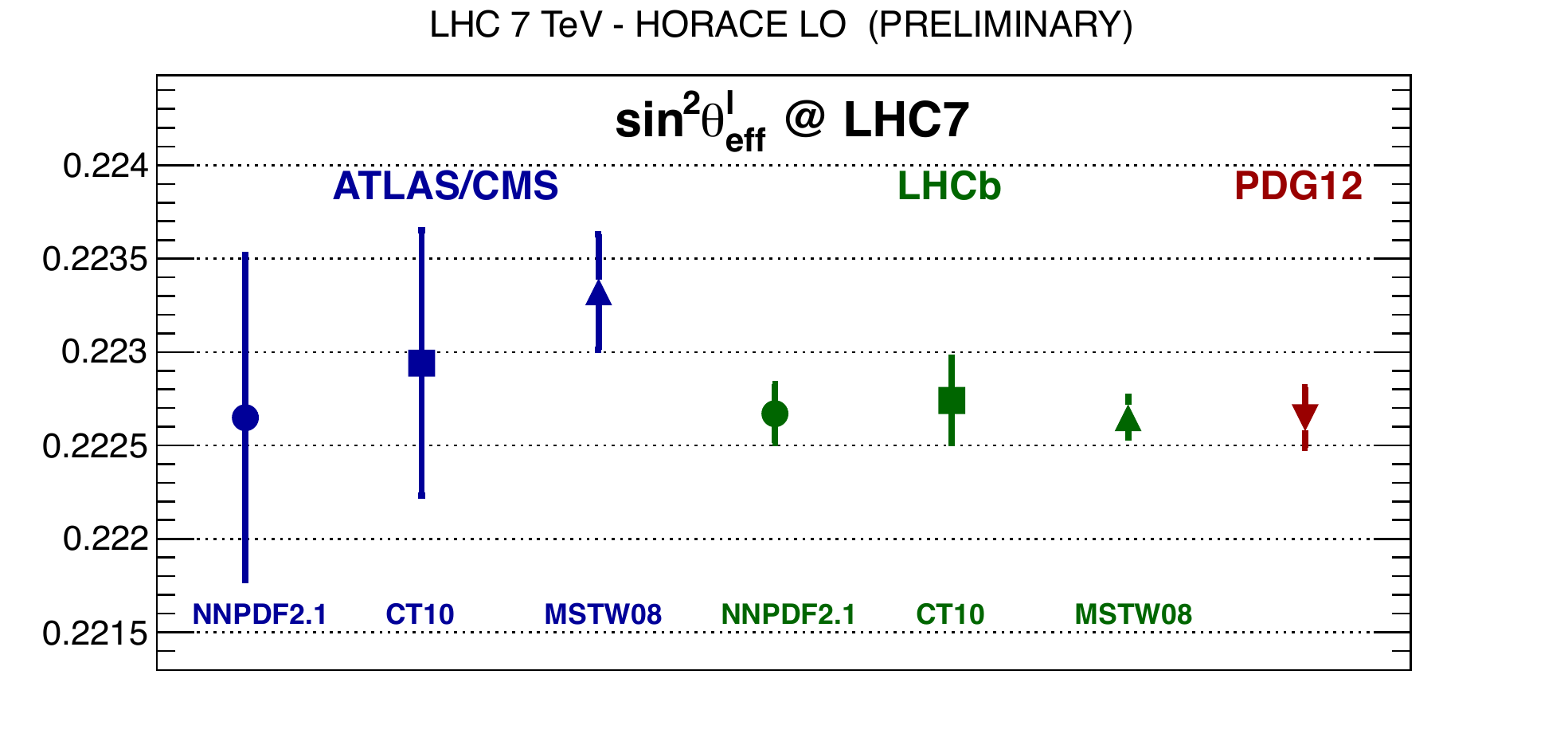}
\caption{\small  PDF uncertainties in the determination of
$\sin^2\theta^l_{\rm eff}$ at the LHC 7 TeV, from \cite{prep2}. The templates have
been computed at LO with the {\tt HORACE} program. We show the results
for NNPDF, MSTW and CT, both for the ATLAS/CMS and the LHCb acceptances.
We also show the uncertainty in the current PDF value,
where the central value has been shifted to match the reference one
used in the pseudo-data generation.}
\label{fig:sin2theta}
\end{figure}

Preliminary results for the PDF uncertainties in 
$\sin^2\theta^l_{\rm eff}$ are shown in Fig.~\ref{fig:sin2theta}.
The templates have been generated with the central NNPDF2.1 set.
We show the results for CT10, MSTW and NNPDF2.1, as well as
the current PDG uncertainty of this parameter. The central value of 
$\sin^2\theta^l_{\rm eff}$ in Fig.~\ref{fig:sin2theta} is not relevant
here.
Our analysis seems to indicate that PDF uncertainties are quite
substantial in the ATLAS/CMS acceptance, though they could be improved
with the help of LHC data. On the other hand, for LHCb
kinematics both the spread between different PDF sets and 
PDF uncertainties are small enough to suggest that
the measurement of $\sin^2\theta^l_{\rm eff}$
would not be limited by PDF uncertainties.
Of course, there are other sources of experimental and theory
systematics that should be taken into account, but at least
our exercise suggest a competitive measurement of 
$\sin^2\theta^l_{\rm eff}$ at the LHC could not be limited by
PDF uncertainties.

While this preliminary study is certainly promising,
a more reliable quantification of PDF uncertainties in
$\sin^2\theta^l_{\rm eff}$ at the LHC requires to
generate templates at
NLO to include the contribution from gluon initiated diagrams.
 Work in this direction
is ongoing.


\providecommand{\href}[2]{#2}\begingroup\raggedright\endgroup

\end{document}